\documentclass{ws-procs9x6}            

\usepackage{wrapfig}              
\usepackage{adjustbox}          

\begin{document}
\title{Induced fission of $^{240}$Pu}

\author{Aurel Bulgac$^*$ and Shi Jin}

\address{Department of Physics, University of Washington, Seattle, WA 98195-1560, USA\\
$^*$E-mail: bulgac@uw.edu}
\author{Piotr Magierski}
\address{ Faculty of Physics, Warsaw University of Technology, ulica Koszykowa 75, 00-662 Warsaw, POLAND}
\author{Kenneth J. Roche}
\address{Pacific Northwest National Laboratory, Richland, WA 99352, USA}
\author{Ionel Stetcu}
\address{Theoretical Division, Los Alamos National Laboratory, Los Alamos, NM 87545, USA}

\begin{abstract}

  We study the fission dynamics of $^{240}$Pu within an implementation
  of the Density Functional Theory (DFT) extended to superfluid
  systems and real-time dynamics. 
  We demonstrate the critical role played by 
  the pairing correlations. 
  The evolution is found to be much slower than
  previously expected  in this fully non-adiabatic treatment of nuclear 
  dynamics, where there are no symmetry restrictions and 
  all collective degrees of freedom (CDOF) are allowed to participate in the dynamics.

\end{abstract}

\keywords{Induced fission, $^{240}$Pu, Time-Dependent Density Functional Theory}

\hspace{0.7cm}

\bodymatter

\begin{wrapfigure}{r}{7cm}
\includegraphics[clip,width=7cm]{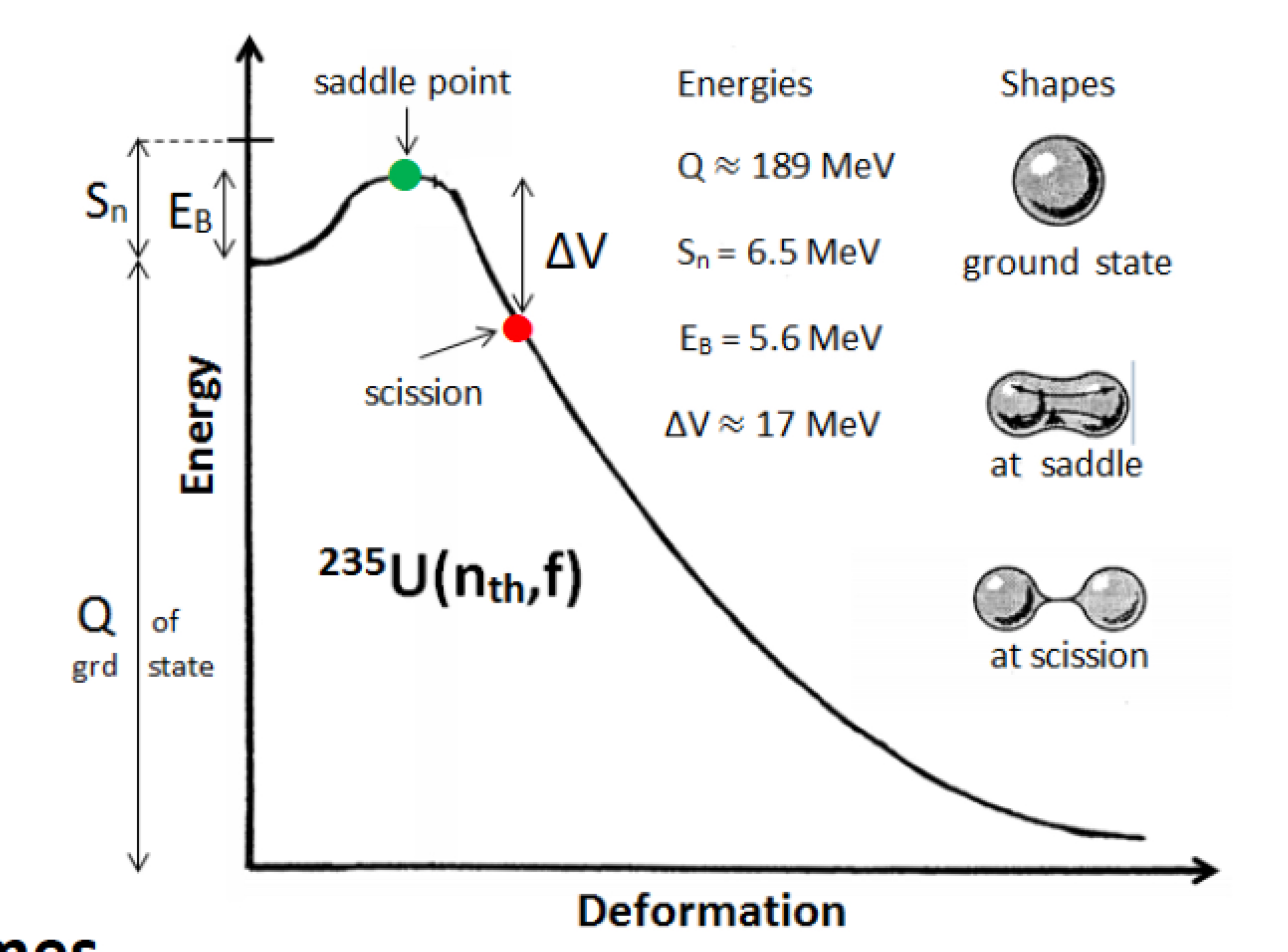}
\includegraphics[clip,width=7cm]{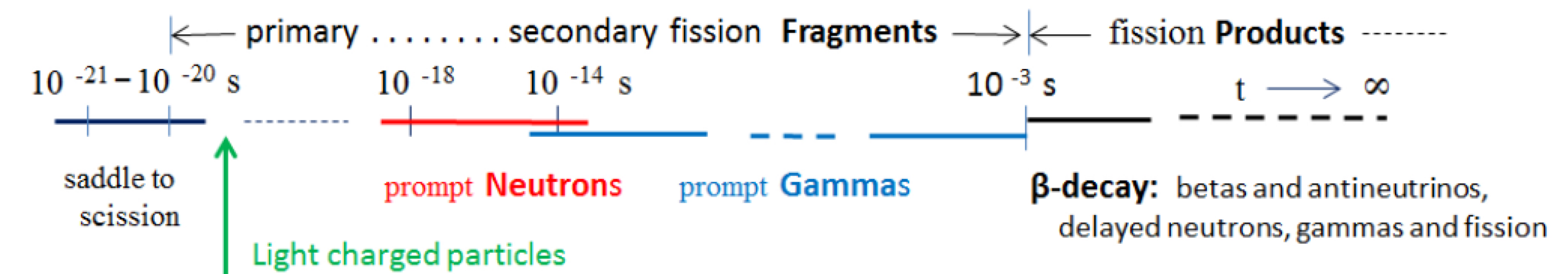}
\caption{ \label{fig:ab1}    Qualitative potential energy of a fissioning nucleus versus deformation and characteristic times of various accompanying processes~\cite{Gonnenwein:2014}.}
\end{wrapfigure}

Nuclear fission, discovered in 1939~\cite{Natur_1939r,Nature_1939r},  is fast approaching the venerable age of 80 years and proves to be one of the most challenging problems
in quantum many-body theory. Nuclear fission is an extremely complex physical phenomenon, starting with the formation of the compound nucleus, the shape evolution until the outer saddle point and eventual slide towards the scission where the fission fragments are formed, accompanied or rather followed by neutron and gamma emissions, followed later on by beta-decay, with times scales of these processes ranging over more than twenty orders of magnitude, see Fig.~\ref{fig:ab1} and Ref.~\cite{Gonnenwein:2014}. 
Likely the most difficult part of this entire process is the descent from the saddle to the scission configuration which over the years has proved quite difficult to define, and the formation of the fission fragments. How a large nucleus with more than 200 nucleons separates into two fragments, how the mass and charge is distributed, how much excitation energy and angular momentum each fragment acquires in this process, how many neutrons and gammas are emitted and at what stage of the fission dynamics, and why and how sometimes even more than two fission fragments are formed? 
Even tough an enormous body of experimental results exists and a large number of phenomenological models have been developed, the present day microscopic results are far from satisfactory~\cite{arxiv_1511r}. 
Nuclear fission is thus unlike another remarkable problem of the quantum many-body problem, namely superconductivity, which since its discovery in 1911~\cite{KNAV_1911r}  was successfully described microscopically in less than 50 years~\cite{PhysRev_1957r}.

\begin{wrapfigure}{r}{5cm}
\includegraphics[clip,width=5cm]{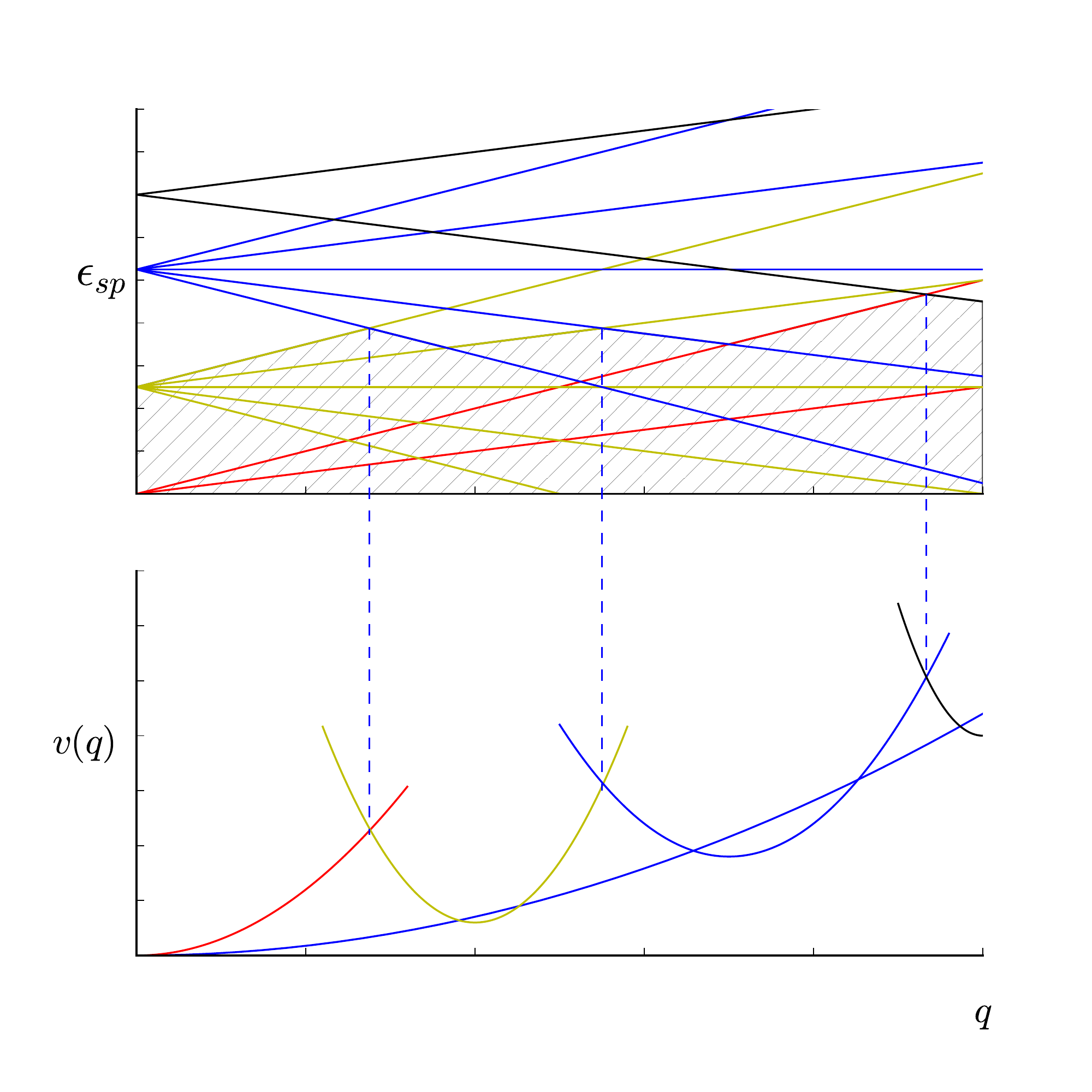}
\caption{ \label{fig:ab2}    The qualitative evolution of the single-particle levels  (upper panel) and of the total nuclear energy (lower panel) as a function of  
nuclear deformation~\cite{PhysRev_1953r, Bertsch:1980}.}
\end{wrapfigure}
Qualitatively it was understood a long time ago that fission can be described qualitatively using a hydrodynamic description of the nucleus as a  charged liquid drop~\cite{Nature_1939r} and that the evolving shape of the fissioning nucleus can be described within a collective model with a deformation potential, see Fig.~\ref{fig:ab1}.  The independent particle model~\cite{Haxel} forced us realize that this smooth deformation potential  actually should have quite a lot of roughness, due to the single-particle level crossings as a function of the nuclear deformation, see Fig.~\ref{fig:ab2} and Ref.~\cite{PhysRev_1953r}. Only the lowest A-levels remain mostly occupied while the nuclear shape evolves, as the nucleus does not heat up significantly and the Fermi surface should retain its spherical overall shape. While the nucleus elongates, the Fermi surface becomes oblate, and it can recover its sphericity only if at a level crossing nucleons from the occupied upward going levels jump to unoccupied downward going levels, see upper panel of Fig.~\ref{fig:ab2}. 
The total energy of the nucleus, which is to a large extent the sum of the occupied single-particle energies, develops cusps at the level crossings. 
It was assumed that the residual interactions between the independent particles provide the mechanism for jumping from one level to another at the crossing, and it was also expected that as result also the deformation potential will become smoother.  Due to Kramers degeneracy each single-particle level is however double-occupied and the only residual interaction capable of providing an effective mechanism to promote simultaneously two particles from one level to another is the pairing 
interaction~\cite{Bertsch:1980}. 

\begin{wrapfigure}{R}{0.6\textwidth}
\includegraphics[clip,width=0.6\textwidth]{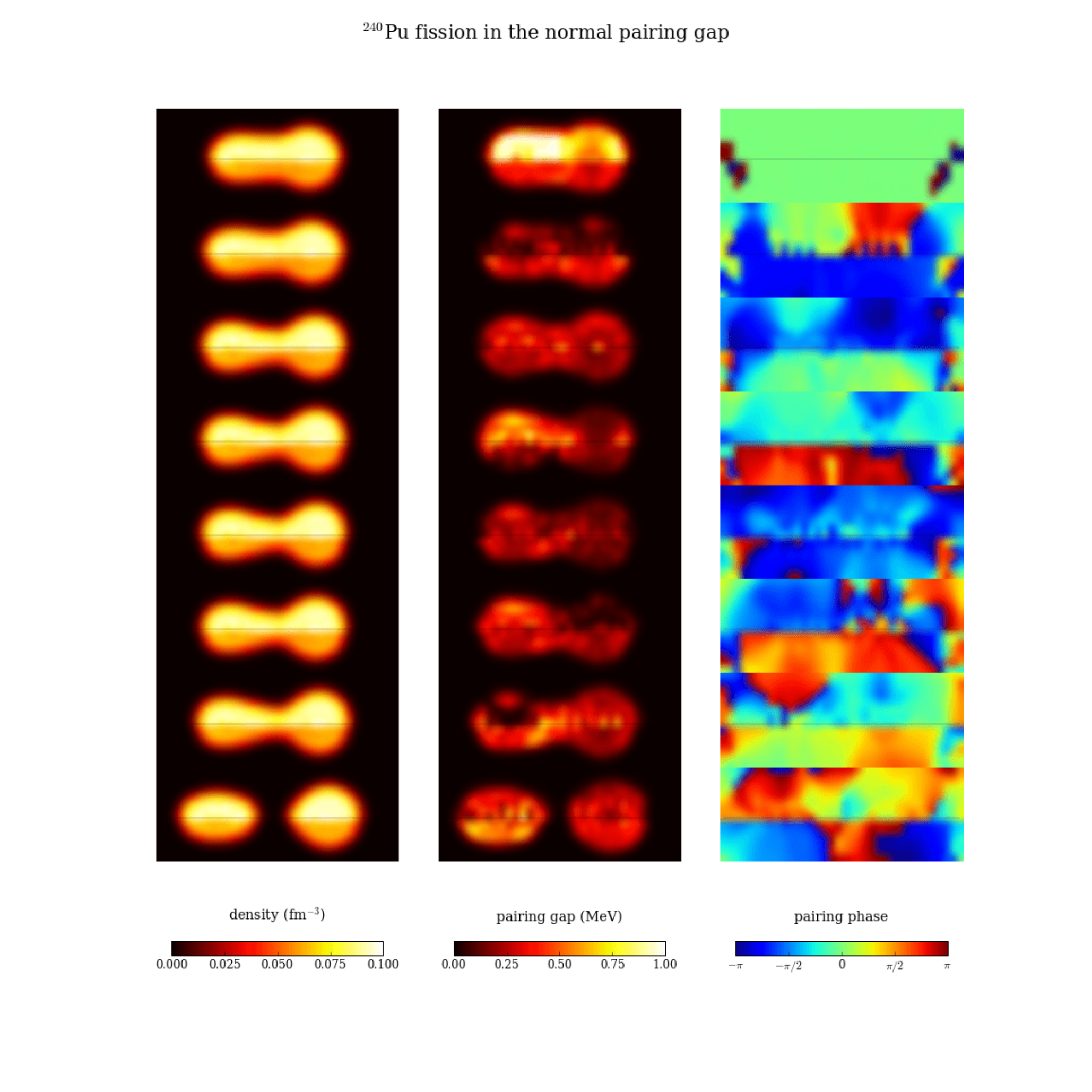}
\caption{ \label{fig:ab3}   Induced fission of $^{240}$Pu with normal pairing strength last about 14,000 fm/c from saddle-to-scission. The columns show sequential frames of the density (first column), magnitude of the pairing field (second column), and the phase of the pairing field (third column). 
In each frame the upper/lower part of each frame shows the neutron/proton density, the magnitude of neutron/proton pairing fields, and of the phase of the pairing field respectively~\cite{Bulgac:2016}.  }
\end{wrapfigure}

This requirement can be understood from a different point of view as well~\cite{PhysRevLett_1997r}. During fission the axial symmetry is typically conserved, and one can expect that the probability distribution of the projections of the angular momenta along the fission axis is also conserved. The initial nucleus has a wider waist than the final fragments and the maximum angular momentum is roughly $p_FR_A$, where $p_F$ is the Fermi momentum and $R_A$ the waist radius. In the final fission products in case of symmetric fission $R_A\rightarrow R_{A/2}=R_A/2^{1/3}$ and thus the maximum angular momentum projection is smaller by a factor of $2^{1/3}$ than in the initial nucleus. A dynamics which will conserve the axial symmetry will not be capable to allow for such a dramatic redistribution of angular momenta of the occupied states. This is the main reason while many attempts to describe fission within a time-dependent Hartree-Fock approach failed so far~\cite{PhysRevC_2014r11}. The pairing interaction is the most effective at coupling nucleon pairs in time reverse quantum states in $(m,-m)\rightarrow(m',-m')$  and it preserves the axial symmetry as well.   Thus a full microscopic treatment of the pairing interactions in a dynamic approach is crucial in describing fission dynamics, see Fig.~\ref{fig:ab3} and Ref.~\cite{Bulgac:2016}. The approach adopted in Ref.~\cite{Bulgac:2016} is based on an extension of the density functional theory to superfluid anytime-dependent phenomena~\cite{ARNPS__2013}. This approach satisfies all expected symmetries of a nuclear Hamiltonian: translational and rotational invariance, Galilean invariance, isospin invariance up to Coulomb effects and proton/neutron mass difference, gauge symmetry, and renormalizability of the theory. 
The static and the time dependent formalism has been confronted with a multitude of theoretical tests and with various experimental data in cold atom physics, nuclear physics, and neutron star crust problems. 

Even though the nuclear energy density functional is not yet known with high enough accuracy and its origin is mostly phenomenological, its basic properties (volume energy, surface energy, symmetry energy, Coulomb energy, spin-orbit interaction) are such that a very large body of observables (masses, charge radii, collective state) are described rather well. For the normal part of the energy density functional we have chosen a well studied parametrization, the SLy4~\cite{NuclPhys_1998}, and for the pairing part we used~\cite{PRL__2003a}. This type of parametrizations of the nuclear energy density functional has met with difficulties when describing spontaneous fission life-times, since for an under-the-barrier process the life-times depend exponentially on the energy density functional parameters. In the case of induced fission, where the entire dynamics occurs in classically allowed regions, inaccuracies of the order of $\cal{O}$(1) MeV in the total deformation potential energy have a relatively small impact on the observables, such as masses and charges of the fission fragments, total kinetic energy and the excitation energies of the fragments. 
The nucleus $^{240}$Pu was prepared in a state close in deformation to the outer fission barrier and an equivalent neutron energy in the reaction $^{239}$Pu(n,f) of about 1.5 MeV. Our goal was not to describe correctly various fission fragments properties, as for many decades the main difficulty the theory was its inability produce fission starting above but near the top of the fission barrier in a real-time approach.The approximate approaches used widely and based on constructing at first a potential energy surface in a collective space of a typically arbitrary dimension between 2 and 5, which was combined with a recipe to calculate also an appropriate inertia tensor in this collective space, even though they might lead to some reasonable predictions, they do not really prove that a truly microscopic theory is at hand.  First of all the choice of collective variables is not rigorous, it is based often on the ability of a specific researcher or group of researchers to solve numerically the problem in the space chosen. It is computationally extremely expensive to construct potential energy surfaces and related inertia tensors in large dimensional spaces. The choice of collective variable is not dictated by a rigorous theory but rather by ``intuition.'' There are also technical difficulties with defining a potential energy surface in a multidimensional space, which is basically a reduction from an infinite dimensional space to a finite dimensional one, a fact well known in mathematics in catastrophe theory even in the case of finite dimensional spaces.  Apart from these rather technical difficulties, there are physics problems, as the introduction of collective degrees of freedom implies an almost exact separation of the degrees of freedom into collective and intrinsic with no coupling between them. This implies that during the evolution the intrinsic degrees of freedom are assumed to remain ``unexcited,'' which is never the case, unless one deals with fully integrable models. There is always a coupling between collective and intrinsics degrees of freedom, this is why fragments emerge excited. This aspect is trivial to put in evidence, one can start with a small number of collective degrees of freedom excited, such as quadrupole and octupole deformation, and let the nucleus evolve freely, only to discover that in an unrestricted dynamics other degrees of freedom are immediately excited with significant amplitudes. This is one of the main reasons why the present ``microscopic'' approaches, based on limited and arbitrarily chosen number of collective degrees of freedom cannot be recognized as a solution of the large amplitude collective nuclear many-body problem. The only viable alternative is to allow all degrees of freedom to be active. Even though this might appear as an insurmountable numerical problem, in fact the problem can be solved with current computers. In a high accuracy simulation of induced fission of $^{240}$Pu we integrated in time numerically 256,000 3D time-dependent coupled non-linear complex partial differential equations on a $25^2\times50$ fm$^3$ spatial lattice for about 320,000 time steps  using 512 GPUs in about 47 hours or using 1602 GPUs in about 24 hours. The lattice constant in this calculation corresponds to a cutoff momentum of $\approx 500$ MeV/c, which is very high and of the same magnitude with the cutoff momenta used in chiral perturbation effective theories of nucleon-nucleon interactions.

\begin{wrapfigure}{R}{0.6\textwidth}
\includegraphics[clip,width=0.6\textwidth]{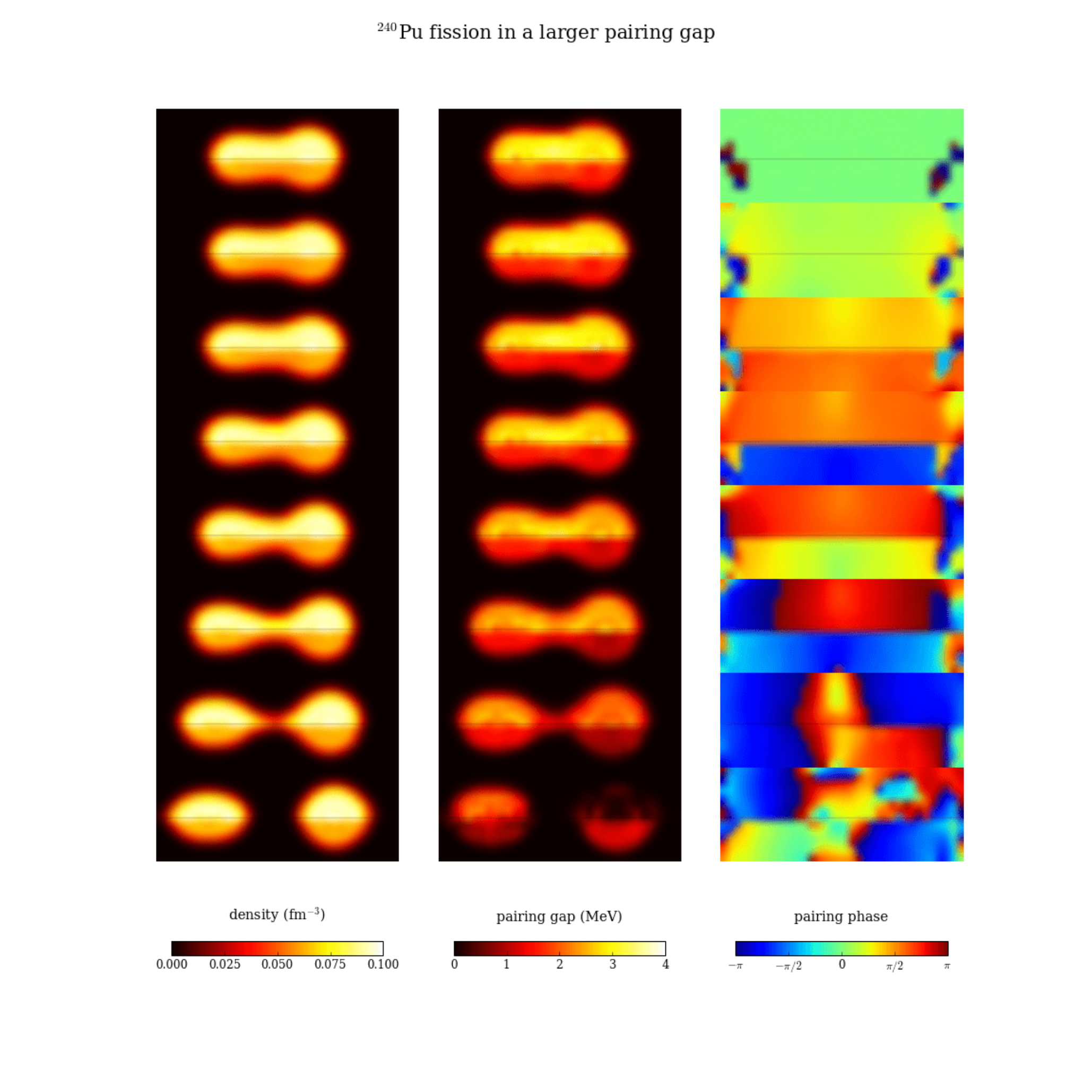}
\caption{ \label{fig:ab4}    Induced fission of $^{240}$Pu with enhanced pairing strength last about 1,400 fm/c from saddle-to-scission, thus about ten times faster than in the case of normal pairing strength.}
\end{wrapfigure}

The outcome of allowing all collective degrees of freedom to be active and to include time and space dependent pairing fields has been remarkable in several ways. The first surprise was that for the first time  an actinide could fission while the dynamics was described with a realistic energy density functional. This could not have happened if pairing correlations would have not been included dynamically, and even if pairing correlations would have been treated approximately at the BCS level either statically or in a time-dependent approach as in Refs.~\cite{PhysRevC_2014r11,Tanimura:2015}.
The second surprise was that the properties of the fission fragments came out very close to the observed ones, even though no effort has been put in trying to obtain them. The physics embodied in the nuclear energy density functional is to a large extent accurate and we attribute this agreement to this fact. The third surprise was that the evolution time from the saddle-to-scission was almost an order of magnitude longer than previously predicted~\cite{PhysRevC_1976r,Bulgac:2015}, 
namely of the order of 10,000 fm/c. It has already been established that in the absence of pairing, or when pairing as a normal strength and is treated in BCS approximation or with frozen initial occupation probabilities, a nucleus will not fission starting at the outer saddle~\cite{Tanimura:2015}, as there is no mechanism to allow for a redistribution of single-particle occupation probabilities. 
In order to prove convincingly the crucial role played by the pairing correlations in the fission dynamics we have increased artificially the strength of the pairing interaction. 
The saddle-to-scission time is then reduced dramatically to about 1,400 fm/c, see Fig.~\ref{fig:ab4}. The pairing field fluctuates both in magnitude and phase at normal pairing strength, Fig.~\ref{fig:ab3}, while these fluctuations are basically absent in case of strong pairing, Fig.~\ref{fig:ab4}, when the dynamics is, as expected, similar to the ideal hydrodynamics~\cite{Bulgac:2015}. The potential energy surface has a lot of ``roughness'' for normal pairing strength, and the slide down of the nucleus is similar to the motion of an electron in the Drude model of electric conductivity, when the electron is kicked out of the direction of the electric field by elastic collisions with the ions, the length of the trajectory is longer, and the average velocity along the direction of the filed is significantly reduced, even though there is no friction. Similarly, the nucleus from saddle-to-scision remains rather cold and only collective degrees of freedom are significantly excited. Pairing, while not being the engine, it provides the essential ``lubricant,'' without which fission is brought to a ``screeching halt.''


\begin{thebibliography}{10}

\bibitem{Natur_1939r} O. Hahn and F. Strassmann, 
             \"Uber den Nachweis und das Verhalten der bei der Bestrahlung des 
              Urans mittels Neutronen entstehenden Erdalkalimetalle,
               Naturwissenschaften {\bf 27}, 11 (1939).
               
\bibitem{Nature_1939r}   L. Meitner and O.R. Frisch, 
              Disintegration of Uranium by Neutron: a New Type of Nuclear Reaction,
              Nature (London), {\bf 143}, 239 (1939).   
              
\bibitem{Gonnenwein:2014} F. G{\" o}nnenwein, Lectures presented at LANL FIESTA Fission School \& Workshop, Sep. 8-12, 2014, Santa Fe, New Mexico, USA, \\ $http://t2.lanl.gov/fiesta2014/$
              
              
\bibitem{arxiv_1511r} N. Schunck and L.M. Robledo, 
              Microscopic Theory of Nuclear Fission: A review,
              Rep. Prog. Phys. {\bf 79}, 116301 (2016).                     
                                
\bibitem{KNAV_1911r} H. Kamerlingh Onnes, 
              Further Experiments with Liquid Helium,
              KNAW, Proceedings, 13 II, 1910-1911, Amsterdam, 1911, pp. 1093-1113.
              
\bibitem{PhysRev_1957r}   J. Bardeen, L. N. Cooper, and J. R. Schrieffer,
              Theory of Superconductivity,
              Phys. Rev. {\bf 108}, 1175 (1957).         
              
\bibitem{Haxel} O.J. Haxel, H.D. Jensen, and H.E. Suess, On
the "magic numbers" in nuclear structure, Phys. Rev. {\bf 75}, 1766 (1949);
                         Maria Goeppert Mayer, On closed shells in nuclei. II, Phys. Rev. {\bf 75}, 1969 (1949).

%
              

                     
%
%
              
%
%
%
                      

\bibitem{PhysRev_1953r}   D.L. Hill and J.A. Wheeler,
              Nuclear Constitution and the Interpretation of Fission Phenomena,
              Phys. Rev. {\bf 89}, 1102 (1953).       
              
\bibitem{Bertsch:1980} G.F. Bertsch, The nuclear density of states in the  space of nuclear shapes, 
 Phys. Lets. {\bf B 95}, 157 (1980);          
              F. Barranco, G.F. Bertsch, R.A. Broglia, and E. Vigezzi,
              Large-Amplitude Motion in Superfluid Fermi Droplets,
              Nucl. Phys. {\bf A512}, 253 (1990);
              G.F. Bertsch,
              Large Amplitude Collective Motion,     
              Nucl. Phys. {\bf A574}, 169c (1994).              
                           
\bibitem{PhysRevLett_1997r} G.F. Bertsch and A. Bulgac,
              Comment on ``Spontaneous Fission: A Kinetic Approach," 
              Phys. Rev. Lett. {\bf 79}, 3539 (1997).          
             
\bibitem{PhysRevC_2014r11} C. Simenel and A.S. Umar,
              Formation and Dynamics of Fission Fragments,
              Phys. Rev. C {\bf 89}, 031601(R) (2014);
              G. Scamps, C. Simenel, and D. Lacroix,
              Superfluid Dynamics of $^{258}$Fm Fission,
              Phys. Rev. C {\bf 92}, 011602(R) (2015).
              
\bibitem{Tanimura:2015}              Y. Tanimura, D. Lacroix, and G. Scamps, 
              Collective Aspects Deduced from the Time-Dependent Microscopic Mean-Field with Pairing:
              Application to the Fission Process,
              Phys. Rev. C {\bf 92}, 034601 (2015);
              P. Goddard, P. Stevenson, and A, Rios,
              Fission Dynamics within Time-Dependent Hartree-Fock: Deformation Induced Fission,
              Phys. Rev. C {\bf  92}, 054610 (2015);     
              P. Goddard, P. Stevenson, and A, Rios,
              Fission Dynamics within Time-Dependent Hartree-Fock: Boost Induced Fission,
              Phys. Rev. C {\bf 93}, 014620 (2016).                            
               
 \bibitem{Bulgac:2016} A. Bulgac, P. Magierski, K.J. Roche, and I. Stetcu,     
               Induced Fission of $^{240}$Pu within a Real-Time  Microscopic Framework, 
               Phys. Rev. Lett. {\bf 116}, 122504 (2016).                                     

\bibitem{ARNPS__2013} A. Bulgac,
             Time-Dependent Density Functional Theory and 
             Real-Time Dynamics of Fermi Superfluids,
              Ann. Rev. Nucl. Part. Sci. {\bf 63}, 97 (2013).      
              
\bibitem{NuclPhys_1998}     E. Chabanat, P. Bonche, P. Haensel, J. Meyer,  R. Schaeffer,
              A Skyrme parametrization from subnuclear to neutron star densities Part II. 
              Nuclei far from stabilities, Nucl. Phys. {\bf A635}, 231 (1998). 
                     
    
\bibitem{PRL__2003a} Y. Yu and A. Bulgac,
              Energy Density Functional Approach to Superfluid Nuclei,
              Phys. Rev. Lett. {\bf 90}, 222501 (2003).                 
          
\bibitem{PhysRevC_1976r} K.T.R. Davies, A.J. Sierk, and J.R. Nix,
              Effect of Viscosity on the Dynamics of Fission,
              Phys. Rev. C {\bf 13}, 2385, (1976);
              J. Blocki, Y. Boneh, J.R. Nix, J. Randrup, M. Robel, A.J. Sierk, W.J. Swiatecki,
             One-body dissipation and the super-viscidity of nuclei,
             Ann. Phys. {\bf  113}, 330 (1978);
              J. Randrup, W.J. Swiatecki,
             One-body dissipation and nuclear dynamics,
             Ann. Phys. {\bf  125}, 193 (1980);
             J.W. Negele, S.E. Koonin, P. M\"{o}ller, J.R. Nix, and A.J. Sierk,
              Dynamics of Induced Fission,
              Phys. Rev. C {\bf 17}, 1098 (1978).   
                          
\bibitem{Bulgac:2015} A. Bulgac, M.M. Forbes, and S. Jin, Nuclear energy density functionals: what do we really know?, arXiv:1506.09195.      
              
              
   
              
             
              
\end{thebibliography}
\end{document}